\documentstyle[12pt]{article}
\begin{document}

\begin{center}

{\large \bf Non-Markovian source term for particle production\\
            by a self-interacting scalar field in the large-$N$ approximation}

\vspace{2 cm}

{\bf Fuad M. Saradzhev}\footnote{e-mail:
fuad\_saradzhev@hotmail.com}

\vspace{5 mm}

{\it Institute of Physics, National Academy of Sciences of
Azerbaijan,\\
H. Javid pr. 33, 370143 Baku, Azerbaijan}

\vspace{1 cm}

{\bf Abstract}

\end{center}

\vspace{5 mm}

The particle production in the self-interacting $N$-component
complex scalar field theory is studied at large $N$. A
non-Markovian source term that includes all higher order
back-reaction and collision effects is derived. The kinetic
amplitudes accounting for the change in the particle number
density caused by collisions are obtained. It is shown that the
production of particles is symmetric in the momentum space. The
problem of renormalization is briefly discussed.

\vspace{5 mm}

PACS: 11.10.z; 05.20.Dd; 11.10.Ef; 98.80.Cq

\vspace{5 mm}

%%%%%%%%%%%%%%%%%%%%%%%%%%%%%%%%%%%%%%%%%%%%%%%%%%%%%%%%%%%%%%%%%%%%%%
\section{Introduction}
%%%%%%%%%%%%%%%%%%%%%%%%%%%%%%%%%%%%%%%%%%%%%%%%%%%%%%%%%%%%%%%%%%%%%%

In recent years an essential progress has been achieved in
establishing a consistent kinetic description of particle
production in quantum field theory from first principles.
Investigations performed for QED with a constant \cite{RM1996} or
time-dependent external electric field \cite{KME1998,S1998} have
shown that a consistent field-theoretical approach leads to a
kinetic equation with a modified source term, providing a
non-Markovian evolution of the distribution function.

Non-Markovian effects in collision terms have been studied  in
connection with relativistic heavy ion collisions \cite{GWR1994},
collective effects in nuclear matter \cite{AYGS1998}, nuclear
fragmentation \cite{CDG1994}, and the damping rates of giant
dipole resonances \cite{FMW1998}. The kinetic description of
transport phenomena in QCD \cite{G1996,B1999} and scalar field
theories \cite{BLL1996} has been also developed. A method to
obtain quantum kinetic equations directly from the underlying
quantum field theory implementing the Schwinger-Keldysh formalism
and a dynamical renormalization group resummation has been
presented in \cite{BVW2000} . This method has been applied to
different scalar and gauge theories, including the
self-interacting ${\varphi}^4$-theory, yielding, however, no
memory effects.

In the present paper, we aim to derive a non-Markovian quantum
kinetic equation for a self-interacting scalar field theory. We
consider the model of a massive $N$-component complex scalar field
in the large-$N$ approximation. Our purpose is to obtain a
non-Markovian source term which incorporates all higher order,
i.e. back-reaction and collisions effects.

In contrast with \cite{BVW2000} , in our paper we follow a kinetic
approach introduced recently to study the decay of $CP$-odd
metastable states and based originally on the evolution operator
technique \cite{B2002} . Herein we generalize this approach to
include higher order effects.

Our plan is as follows. In Section 2 we introduce the model and
perform the quantization of the scalar field fluctuations within
the mean-field approximation. We construct the fluctuations
Hamiltonian and identify its diagonalized part. In Section 3 we
study at large $N$ the effects of the remaining nondiagonalized
interactions in the particle production. We derive a quantum
kinetic equation governing the production process and discuss its
renormalization. We conclude with summary in Section~4.

%%%%%%%%%%%%%%%%%%%%%%%%%%%%%%%%%%%%%%%%%%%%%%%%%%%%%%%%%%%%%%%%%%%%%%
\section{Quantization}
%%%%%%%%%%%%%%%%%%%%%%%%%%%%%%%%%%%%%%%%%%%%%%%%%%%%%%%%%%%%%%%%%%%%%%

The Lagrangian density of our model is
\begin{equation}
{\cal L}= \sum_{j=1}^{N} \Big[ ({\partial}_{\mu} {\varphi}_{j}^{\dagger})
({\partial}^{\mu} {\varphi}_{j})
- m^2 {\varphi}_{j}^{\dagger} {\varphi}_{j} \Big]
- V({\varphi}),
\label{1}
\end{equation}
where $m$ is a bare mass. The model is defined in a finite volume
$L^3$, $-L/2 \leq x_i \leq L/2$, $i=1,2,3$, and the scalar field
fulfils the periodic boundary conditions. In the continuum limit
$\frac {1}{L^3} \sum_{\vec k}$ , where the summation is over
discrete momenta $\vec{k} = \frac{2\pi}{L} \vec{n}$,
$(\vec{n})=(n_1,n_2,n_3)$, goes to $\int
\frac{d^3\vec{k}}{(2{\pi})^3}$.

The potential is taken as
\begin{equation}\label{2}
V({\varphi})=\frac{\lambda}{2} \Big( \sum_{j}^{N}
{\varphi}_{j}^{\dagger} {\varphi}_{j} \Big)^2,
\hspace{1 cm}
{\lambda}>0,
\end{equation}
and includes self-interaction for each component as well as interaction
of different components. From (\ref{1}) we obtain the Klein-Gordon type
equation of motion
\begin{equation}
(\Box + m^2) {\varphi}_{j} = J_{j}
\label{3}
\end{equation}
with the current
\begin{equation}
J_j \equiv - \frac{{\partial}V}{{\partial}{\varphi}_{j}^{\dagger}}
= - {\lambda} \Big( \sum_{l=1}^{N} {\varphi}_{l}^{\dagger}
{\varphi}_{l} \Big) {\varphi}_j. \label{4}
\end{equation}
%

%%%%%%%%%%%%%%%%%%%%%%%%%%%%%%%%%%%%%%%%%%%%%%%%%%%%%%%%%%%%%%%%%%%%
\subsection{Mean-field approximation}
%%%%%%%%%%%%%%%%%%%%%%%%%%%%%%%%%%%%%%%%%%%%%%%%%%%%%%%%%%%%%%%%%%%%%

Following the mean-field approximation, we usually decompose
${\varphi}_{j}(\vec{x},t)$ into its vacuum mean value
${\phi}_{j}(t)=\langle {\varphi}_{j}(\vec{x},t) \rangle$ and
fluctuations ${\chi}_{j}$
\begin{equation}
{\varphi}_{j}(\vec{x},t) = {\phi}_{j}(t) + {\chi}_{j}(\vec{x},t)
\label{5}
\end{equation}
with $\langle {\chi}_{j}(\vec{x},t) \rangle =0$.
The mean field is treated as a classical background field
defined with respect to the in-vacuum
$|0 \rangle$ as
\begin{equation}
{\phi}_{j}(t) \equiv
\frac{1}{L^3} \int d^3x \langle 0| {\varphi}_{j}(\vec{x},t) |0 \rangle,
\label{6}
\end{equation}
so that in the limit $t \to -\infty$  ${\phi}_{j}(t) \to 0$,
while the fluctuations are quantized and take place at all times.

The background ${\phi}$-fields are in general complex and
contribute to the charge of the system. By a global $U(1)$ gauge
transformation the phases of the ${\phi}$-fields can be put equal
to zero at a fixed time. If the vanishing of the phases is
compatible with equations of motion, it is possible to make them
zero at all times as well. Otherwise, the background phases become
dynamical variables coupled to the fluctuations in all orders.
This results in new, non-diagonal terms in the Hamiltonian and
therefore in new interactions.

 To simplify the problem, it is
convenient to choose the background neutral. Let us introduce the
real fields
\begin{equation}
{\phi}_{1j} \equiv \frac{1}{2} ( {\phi}_j + {\phi}_j^{\star} ),
\label{7}
\end{equation}
\begin{equation}
{\phi}_{2j} \equiv \frac{1}{2i} ( {\phi}_j - {\phi}_j^{\star} ),
\label{8}
\end{equation}
and rewrite the decomposition (\ref{5}) as
\begin{equation}
{\varphi}_j(\vec{x},t) = {\phi}_{1j}(t) +
\overline{\chi}_j(\vec{x},t),
\label{9}
\end{equation}
where $\overline{\chi}_j \equiv {\chi}_j + i {\phi}_{2j}$. The
vacuum mean value of the "new" fluctuations is non-vanishing,
$\langle \overline{\chi}_j \rangle = i {\phi}_{2j}$.

Using Eq.(\ref{9}) provides
the following decomposition for the current
\begin{equation}
\label{10} J_j(\varphi) = J_j({\phi}_{1}) - {\mu}_0^2
\overline{\chi}_j - {\mu}_j \sum_{l=1}^{N} {\mu}_l (
\overline{\chi}_l + \overline{\chi}_l^{\dagger} )+
\overline{J}_j({\phi};\overline{\chi}),
\end{equation}
where
\begin{equation}
{\mu}_0^2 \equiv \sum_{j=1}^{N} {\mu}_j^2, \label{11}
\end{equation}
\begin{equation}
{\mu}_j \equiv \sqrt{\lambda} {\phi}_{1j},
\label{12}
\end{equation}
while $\overline{J}_j({\phi};\overline{\chi})$ includes terms of second
and higher orders in fluctuations,
\begin{equation}
\overline{J}_j({\phi};\overline{\chi})
 \equiv - \sqrt{\lambda} {\mu}_j \sum_{l=1}^{N}
 \overline{\chi}_l^{\dagger} \overline{\chi}_l
 - \sqrt{\lambda} \overline{\chi}_j \sum_{l=1}^{N} {\mu}_l
 (\overline{\chi}_l + \overline{\chi}_l^{\dagger})
 + J_j(\overline{\chi}).
\label{13}
\end{equation}
Substituting Eq.(\ref{9}) also into (\ref{3}) and taking the mean value
$\langle . . . \rangle$ yields the vacuum mean field equations
\begin{equation}
\ddot{\phi}_{1j} + (m^2+{\mu}_0^2) {\phi}_{1j}= {\rm Re} \langle
\overline{J}_j({\phi};\overline{\chi}) \rangle, \label{14}
\end{equation}
\begin{equation}
\ddot{\phi}_{2j} + (m^2+{\mu}_0^2) {\phi}_{2j}=
{\rm Im} \langle \overline{J}_j({\phi};\overline{\chi}) \rangle.
\label{15}
\end{equation}
As seen from (\ref{15}), only if higher orders are neglected can
the ${\phi}$-fields be taken neutral. Assuming that
${\phi}_{2j}=0$ leads to the condition
\begin{equation}
{\rm Im} \langle \overline{J}_j({\phi};\overline{\chi}) \rangle
({\phi}_2=0) =0,
\label{16}
\end{equation}
which is not valid in general if the fluctuations are charged.

Eq.(\ref{3}) in concert with Eqs.(\ref{14})-(\ref{15}) provides
also the equations of motion for the fluctuations. Introducing instead
of $\overline{\chi}_j$,$\overline{\chi}_j^{\dagger}$ the hermitian
fields
\begin{equation}
\overline{\chi}_{1j} \equiv \frac{1}{\sqrt{2}}
(\overline{\chi}_j + \overline{\chi}_j^{\dagger}),
\label{17}
\end{equation}
\begin{equation}
\overline{\chi}_{2j} \equiv \frac{1}{i\sqrt{2}}
(\overline{\chi}_j - \overline{\chi}_j^{\dagger}),
\label{18}
\end{equation}
with $\langle \overline{\chi}_{1j} \rangle =0$,
$\langle \overline{\chi}_{2j} \rangle={\phi}_{2j}$, yields
\begin{equation}
\Big( \Box + m_{{\alpha}j}^2 \Big) \overline{\chi}_{{\alpha}j} =
F_{{\alpha}j}, \label{19}
\end{equation}
$({\alpha}=1,2)$, where
\begin{equation}
m_{1j}^2 \equiv m^2 + {\mu}_0^2 + 2{\mu}_j^2, \label{20}
\end{equation}
\begin{equation}
m_{2j}^2 \equiv m^2 + {\mu}_0^2 \label{21}
\end{equation}
are effective masses, and
\begin{equation}
F_{1j} \equiv \overline{J}_{1j} - \langle \overline{J}_{1j}
\rangle - 2{\mu}_j \sum_{\stackrel{l=1}{l \neq j}}^{N} {\mu}_l
\overline{\chi}_{1l}, \label{22}
\end{equation}
\begin{equation}
F_{2j} \equiv \overline{J}_{2j}, \label{23}
\end{equation}
the hermitian currents $\overline{J}_{{\alpha}j}$ being defined
instead of $\overline{J}_j$, $\overline{J}_j^{\dagger}$ in the way
analogous to (\ref{17})-(\ref{18}).

Since the effective masses squared are always positive, the system
evolves only in the non-tachyonic regime ( for the kinetic
approach to the tachyonic regime, see \cite{S2003} ). For $N=1$,
the effective masses squared  take the form $m_1^2=m^2 +
3{\lambda}{\phi}_{1}^2$, $m_2^2=m^2 + {\lambda}{\phi}_{1}^2$ in
agreement with \cite{K1989}.

Rewriting (\ref{19}) for the Fourier components
$\overline{\chi}_{{\alpha}j}(\vec{k},t)$, $\alpha=1,2$, we obtain
a Mathieu type equation
\begin{equation}
\ddot{\overline{\chi}}_{{\alpha}j}(\vec{k},t) +
{\omega}_{{\alpha}j}^2(\vec{k},t) \overline{\chi}_{{\alpha}j}
(\vec{k},t) = F_{{\alpha}j}(\vec{k},t), \label{24}
\end{equation}
where $F_{{\alpha}j}(\vec{k},t)$ is the Fourier transform of
$F_{{\alpha}j}(\vec{x},t)$, and
\begin{equation}
{\omega}_{{\alpha}j}^2 \equiv {\vec{k}}^2 + m_{{\alpha}j}^2
\label{25}
\end{equation}
are the time-dependent frequencies squared of the fluctuations.
In the in-limit, ${\omega}_{{\alpha}j}$ become
\begin{equation}
\lim_{t \to -\infty} {\omega}_{{\alpha}j}(\vec{k},t) \equiv
{\omega}^{0}(\vec{k}) =\sqrt{(\vec{k})^2+m^2}. \label{26}
\end{equation}
Eqs. (\ref{14}),(\ref{15}) and (\ref{24}) are self-consistently
coupled and include all higher-order effects. The vacuum mean
fields modify the equations for the fluctuations via the time
dependent frequencies, while the fluctuations react back on the
${\phi}$-fields via the source terms ${\rm Re}\langle
\overline{J}_j \rangle$ and ${\rm Im}\langle \overline{J}_j
\rangle$ in Eqs.(\ref{14}),(\ref{15}) and on the fluctuations
themselves via the term $F_{{\alpha}j}(\vec{k},t)$ in
Eq.(\ref{24}).

%%%%%%%%%%%%%%%%%%%%%%%%%%%%%%%%%%%%%%%%%%%%%%%%%%%%%%%%%%%%%%%%%%%%%%%
\subsection{Hamiltonian}
%%%%%%%%%%%%%%%%%%%%%%%%%%%%%%%%%%%%%%%%%%%%%%%%%%%%%%%%%%%%%%%%%%%%%%%

The fluctuations act on each other in two ways: directly, when higher
orders are included, and via the ${\phi}$-fields. The potential
(\ref{2}) is decomposed as
\begin{equation}
V({\varphi}) = V({\phi}_1) + \sqrt{2} {\mu}_0^2 \sum_{j=1}^{N}
{\phi}_{1j} \overline{\chi}_{1j} + \frac{{\mu}_0^2}{2}
\sum_{j=1}^{N} ( \overline{\chi}_{1j}^2 + \overline{\chi}_{2j}^2 )
+ \sum_{(j,l)=1}^{N} {\mu}_j {\mu}_l \overline{\chi}_{1j}
\overline{\chi}_{1l} + \overline{V}({\phi};\overline{\chi}),
\label{27}
\end{equation}
where $\overline{V}({\phi};\overline{\chi})$ contains third
and fourth orders in fluctuations,
\begin{equation}
\overline{V}({\phi};\overline{\chi}) = - \sum_{j=1}^{N}
{\phi}_{1j} \Big( J_{j}(\overline{\chi}) +
J_{j}^{\dagger}(\overline{\chi}) \Big) + V(\overline{\chi}).
\label{28}
\end{equation}
With the decomposition (\ref{27}), we deduce from (\ref{1}) the
Hamiltonian density governing the dynamics of the fluctuations
\begin{equation}
{\cal H}_{\chi}= \frac{1}{2} \sum_{{\alpha}=1}^{2} \sum_{j=1}^{N}
\Big( \overline{\pi}_{{\alpha}j}^2 + (\vec{\nabla}
\overline{\chi}_{{\alpha}j} )^2 + m_{{\alpha}j}^2
\overline{\chi}_{{\alpha}j}^2 \Big) + \sum_{\stackrel{(j,l)=1}{j
\neq l}}^{N} {\mu}_j {\mu}_l \overline{\chi}_{1j}
\overline{\chi}_{1l} + \overline{\cal H}_{\chi}, \label{29}
\end{equation}
where $\overline{\pi}_{{\alpha}j}$ are the momenta canonically
conjugate to $\overline{\chi}_{{\alpha}j}$, and
\begin{equation}
\overline{\cal H}_{\chi} \equiv \overline{V}({\phi}_1;
\overline{\chi}) + \sum_{j=1}^{N} \langle \overline{J}_{1j}
\rangle \overline{\chi}_{1j}. \label{30}
\end{equation}
The expression (\ref{29}) is not diagonal even in the second order
in fluctuations. The non-diagonal terms include interactions of
fields of different values of $j$.

In terms of the Fourier components $\overline{\chi}_{{\alpha}j}
(\vec{k},t)$, $\overline{\pi}_{{\alpha}j}(\vec{k},t)$, the
fluctuations Hamiltonian takes the form
\begin{equation}
H_{\chi} = \int d^3x {\cal H}_{\chi} = \sum_{j=1}^{N} H_j +
\sum_{\stackrel{(j,l)=1}{j \neq l}}^{N} H_{jl} + L^{3/2}
\overline{\cal H}_{\chi} (\vec{k}=0,t), \label{31}
\end{equation}
where $\overline{\cal H}_{\chi}(\vec{k},t)$ is the Fourier
transform of the potential $\overline{\cal H}_{\chi}(\vec{x},t)$,
and
\begin{equation}
H_j \equiv \frac{1}{2} \sum_{{\alpha}=1}^{2} \sum_{\vec{k}} \Big(
\overline{\pi}_{{\alpha}j}^{\dagger}(\vec{k},t)
\overline{\pi}_{{\alpha}j}(\vec{k},t) +
{\omega}_{{\alpha}j}^2(\vec{k},t) \overline{\chi}_{{\alpha}j}
^{\dagger}(\vec{k},t) \overline{\chi}_{{\alpha}j}(\vec{k},t)
\Big), \label{32}
\end{equation}
\begin{equation}
H_{jl} \equiv {\mu}_j {\mu}_l \sum_{\vec{k}}
\overline{\chi}_{1j}(\vec{k},t) \overline{\chi}_{1l}^{\dagger}
(\vec{k},t). \label{33}
\end{equation}
The Hamiltonian equations of motion are
\begin{eqnarray}
\dot{\overline{\chi}}_{{\alpha}j}(\vec{k},t) & = &
\overline{\pi}_{{\alpha}j}^{\dagger}(\vec{k},t),\\
\dot{\overline{\pi}}_{{\alpha}j}(\vec{k},t) & = & -
{\omega}_{{\alpha}j}^2(\vec{k},t) \overline{\chi}_{{\alpha}j}
^{\dagger}(\vec{k},t) + F^{\dagger}_{{\alpha}j}(\vec{k},t),
\label{34-35}
\end{eqnarray}
in agreement with (\ref{24}).

By the transition to the new fields (\ref{17})-(\ref{18}), we have
achieved a partial diagonalization of the Hamiltonian in the
second order in fluctuations. Its complete diagonalization,
including all higher order, i.e. interaction terms, is a very
complicated problem. However, there are different approximation
schemes which allow one to diagonalize a "physically" important
part of the interaction, while the remaining part is treated as a
small perturbation.

Let us assume that, by using one of these schemes, we succeeded in
diagonalising some of the interaction terms in our model. We
assume next that a new diagonalized part of the Hamiltonian looks
like the one in Eq.(\ref{32}), the frequencies of fluctuations
being only modified. Namely, ${\omega}_{{\alpha}j}^2(\vec{k},t)$
is replaced by
\begin{equation}
\tilde{\omega}_{{\alpha}j}^2(\vec{k},t) \equiv
{\omega}_{{\alpha}j}^2(\vec{k},t) + M_{{\alpha}j}^2(\vec{k},t),
\label{36}
\end{equation}
where $M_{{\alpha}j}^2(\vec{k},t)$ represents the higher order
effects contribution. The explicit expression for
$M_{{\alpha}j}^2(\vec{k},t)$ depends on the approximation scheme
under consideration. In the next section, we will calculate
$M_{{\alpha}j}^2(\vec{k},t)$ in the large-$N$ approximation.

For the Hermitian fluctuations, we introduce then the annihilation
and creation operators $d_{{\alpha}j}(\vec{k},t)$ and
$d_{{\alpha}j}^{\dagger}(\vec{k},t)$ in a standard way by
\begin{equation}
\overline{\chi}_{{\alpha}j}(\vec{k},t) = {\Gamma}_{{\alpha}j}
(\vec{k},t) d_{{\alpha}j}(\vec{k},t) +
{\Gamma}_{{\alpha}j}^{\star} (\vec{k},t)
d_{{\alpha}j}^{\dagger}(-\vec{k},t) \label{37}
\end{equation}
and
\begin{equation}
\overline{\pi}_{{\alpha}j}(\vec{k},t) = -i
\tilde{\omega}_{{\alpha}j} (\vec{k},t) \Big[
{\Gamma}_{{\alpha}j}(\vec{k},t) d_{{\alpha}j}(-\vec{k},t) -
{\Gamma}_{{\alpha}j}^{\star} (\vec{k},t)
d_{{\alpha}j}^{\dagger}(\vec{k},t) \Big], \label{38}
\end{equation}
where
\begin{equation}
{\Gamma}_{{\alpha}j}(\vec{k},t) = \frac{1}{\sqrt{
2\tilde{\omega}_{{\alpha}j}(\vec{k},t) }} \exp\{- i
{\Theta}_{{\alpha}j} (\tilde{\omega}_{{\alpha}j},t) \} \label{39}
\end{equation}
and ${\Theta}_{{\alpha}j}(\tilde{\omega}_{{\alpha}j},t)$ are
arbitrary phases. In the case when higher order effects are
omitted , i.e. $M_{{\alpha}j}^2=0$, the phases
${\Theta}_{{\alpha}j}({\omega}_{{\alpha}j},t)$  take the form
${\omega}_k^{(0)}t$ in the in-limit, and the operators
$\tilde{\chi}_{{\alpha}j}(\vec{k},t)$,
$\tilde{\pi}_{{\alpha}j}(\vec{k},t)$ can be connected with the
corresponding in-field ones by making use of the evolution
operator \cite{B2002}.

The ansatz (\ref{37}) ( and (\ref{38}) ) has the same structure as
the free field theory one except the frequencies
$\tilde{\omega}_{{\alpha}j} (\vec{k},t)$ replace
${\omega}^{(0)}(\vec{k})$, and the time-independent annihilation
and creation operators are replaced here by the time-dependent
ones $d_{{\alpha}j}(\vec{k},t)$,
$d_{{\alpha}j}^{\dagger}(\vec{k},t)$. The Hamiltonians $H_j$
become
\begin{equation}
H_j = \frac{1}{2} \sum_{{\alpha}=1}^{2} \sum_{\vec{k}}
\tilde{\omega}_{{\alpha}j}(\vec{k},t) \Big(
d_{{\alpha}j}^{\dagger}(\vec{k},t) d_{{\alpha}j}(\vec{k},t) +
d_{{\alpha}j}(\vec{k},t) d_{{\alpha}j}^{\dagger}(\vec{k},t) \Big).
\label{40}
\end{equation}
%

%%%%%%%%%%%%%%%%%%%%%%%%%%%%%%%%%%%%%%%%%%%%%%%%%%%%%%%%%%%%%%%%%%%
\section{Kinetic equations}
%%%%%%%%%%%%%%%%%%%%%%%%%%%%%%%%%%%%%%%%%%%%%%%%%%%%%%%%%%%%%%%%%%%

There are $2N$ types of neutral particles in our model which are
related to $2N$ types of the fluctuations $\overline{\chi}_{{\alpha}j}$.
The numbers of these particles of momentum $\vec{k}$ at time $t$
are given by the occupation number densities
\begin{equation}
{\cal N}_{{\alpha}j} (\vec{k},t) \equiv \langle
0|d_{{\alpha}j}^{\dagger} (\vec{k},t) d_{{\alpha}j}(\vec{k},t) |0
\rangle. \label{41}
\end{equation}
In the limit $t \to - \infty$, these densities vanish because
the initial state is assumed to be "empty", i.e. without
particles.

In general, ${\cal N}_{{\alpha}j}(\vec{k},t)$ and
${\cal N}_{{\alpha}j}(-\vec{k},t)$ are not equal.
Therefore it is convenient to introduce
\begin{equation}
{\cal N}_{{\alpha}j,{\pm}}(\vec{k},t) \equiv \frac{1}{2} \Big(
{\cal N}_{{\alpha}j}(\vec{k},t) \pm {\cal
N}_{{\alpha}j}(-\vec{k},t) \Big), \label{42}
\end{equation}
where ${\cal N}_{{\alpha}j,+}(\vec{k},t)$ is the particle number
averaged over the directions $\vec{k}$ and $(-\vec{k})$,
while ${\cal N}_{{\alpha}j,-}(\vec{k},t)$ measures the degree
of asymmetry.

We consider first the time evolution of ${\cal
N}_{{\alpha}j,-}(\vec{k},t)$. Using the relations
\begin{equation}
d_{{\alpha}j}(\vec{k},t) =
\frac{1}{2{\Gamma}_{{\alpha}j}(\vec{k},t)} \Big(
\overline{\chi}_{{\alpha}j}(\vec{k},t) +
\frac{i}{\tilde{\omega}_{{\alpha}j}(\vec{k},t)}
\overline{\pi}_{{\alpha}j}^{\dagger}(\vec{k},t) \Big) \label{43}
\end{equation}
and the Hamiltonian equations of motion $(34)-(35)$, we obtain
\begin{equation}
\dot{\cal N}_{{\alpha}j,-}(\vec{k},t) = - {\rm Im} \langle 0|
\overline{\chi}_{{\alpha}j}^{\dagger} (\vec{k},t)
F_{{\alpha}j}(\vec{k},t) |0 \rangle. \label{44}
\end{equation}
If we omit higher-order effects and take $\overline{J}_{{\alpha}j}
=0$, then
\begin{equation}
\dot{\cal N}_{{\alpha}j,-}(\vec{k},t) = 2{\mu}_j
{\delta}_{{\alpha}1} \sum_{\stackrel{l=1}{l \neq j}}^{N} {\mu}_l
{\rm Im} \langle 0| \overline{\chi}_{1j}^{\dagger} (\vec{k},t)
\overline{\chi}_{1l}(\vec{k},t) |0 \rangle, \label{45}
\end{equation}
i.e. the time evolution of the asymmetry in the production of
particles by the field of a fixed value of $j$ is determined by
its interaction with other fields. For $N=1$, the density ${\cal
N}_{{\alpha}1,-} (\vec{k},t)$ is conserved, $\dot{\cal
N}_{{\alpha}1,-}(\vec{k},t)=0$, so if we take ${\cal
N}_{{\alpha}1,-}(\vec{k},t_0)=0$, where $t_0$ is a moment of time
at which the particle production starts, then the asymmetry does
not appear at any time $t$ later.

Taking next the time derivative of ${\cal N}_{{\alpha}j,+}(\vec{k},t)$,
yields the evolution equation
\begin{equation}
\dot{\cal N}_{{\alpha}j,+}(\vec{k},t) =
\frac{\dot{\tilde{\omega}}_{{\alpha}j}}{\tilde{\omega}_{{\alpha}j}}
{\rm Re}\Big[ e^{-2i{\Theta}_{{\alpha}j}} C_{{\alpha}j}(\vec{k},t)
\Big] + {\Delta}_{{\alpha}j}^{(1)}(\vec{k},t), \label{46}
\end{equation}
where we have defined the time-dependent pair correlation functions
\begin{equation}
C_{{\alpha}j}(\vec{k},t) \equiv \langle 0|
d_{{\alpha}j}(-\vec{k},t) d_{{\alpha}j}(\vec{k},t) |0 \rangle
\label{47}
\end{equation}
and
\begin{equation}
{\Delta}_{{\alpha}j}^{(1)} \equiv
\frac{1}{\tilde{\omega}_{{\alpha}j}} {\rm Re} \langle 0|
\overline{\pi}_{{\alpha}j}(\vec{k},t)
\tilde{F}_{{\alpha}j}(\vec{k},t) |0 \rangle \label{48}
\end{equation}
with
\begin{equation}
\tilde{F}_{{\alpha}j}(\vec{k},t) \equiv F_{{\alpha}j}(\vec{k},t) +
M_{{\alpha}j}^2(\vec{k},t) \overline{\chi}_{{\alpha}j}(\vec{k},t).
\label{49}
\end{equation}
The pair correlation functions $C_{{\alpha}j}(\vec{k},t)$ can be
shown to obey the equations
\begin{equation}
\dot{C}_{{\alpha}j}(\vec{k},t) - 2i (\dot{\Theta}_{{\alpha}j} -
\tilde{\omega}_{{\alpha}j}) C_{{\alpha}j}(\vec{k},t) =\Big[
\frac{\dot{\tilde{\omega}}_{{\alpha}j}}{2\tilde{\omega}_{{\alpha}j}}
\Big( 1+2{\cal N}_{{{\alpha}j},+}(\vec{k},t) \Big) + \Big(
-{\Delta}_{{\alpha}j}^{(1)}(\vec{k},t) +i
{\Delta}_{{\alpha}j}^{(2)}(\vec{k},t) \Big) \Big]
e^{2i{\Theta}_{{\alpha}j}}, \label{50}
\end{equation}
where
\begin{equation}
{\Delta}_{{\alpha}j}^{(2)}(\vec{k},t) \equiv {\rm Re} \langle 0|
\overline{\chi}_{{\alpha}j}^{\dagger}(\vec{k},t)
\tilde{F}_{{\alpha}j}(\vec{k},t) |0 \rangle. \label{51}
\end{equation}

As seen from Eqs.(\ref{46}) and (\ref{50}), the higher order
effects contribute to the evolution equations via the frequencies
$\tilde{\omega}_{{\alpha}j}$ and via the functions
${\Delta}_{{\alpha}j}^{(1)}(\vec{k},t)$ and
${\Delta}_{{\alpha}j}^{(2)}(\vec{k},t)$. These functions include
higher order correlation functions, namely, the vacuum expectation
values of the products of three and four field operators.

%%%%%%%%%%%%%%%%%%%%%%%%%%%%%%%%%%%%%%%%%%%%%%%%%%%%%%%%%%%%%%%%%%%
\subsection{Large-$N$ approximation}
%%%%%%%%%%%%%%%%%%%%%%%%%%%%%%%%%%%%%%%%%%%%%%%%%%%%%%%%%%%%%%%%%%%

To calculate ${\Delta}^{(1)}_{{\alpha}j}(\vec{k},t)$,
${\Delta}^{(2)}_{{\alpha}j}(\vec{k},t)$, we use the large-$N$
approximation [15-18]. At large $N$, the theory , which
consists of an infinite number of coupled correlation function
equations, reduces to the solution of coupled equations for the
one- and two-point functions. For simplicity, we consider the case
of spatially homogeneous vacuum configurations and assume that
$\langle 0| \overline{\chi}_{2j}(\vec{x},t) |0 \rangle$ is a
function only of time, then
\begin{equation}
\langle 0| \overline{\chi}_{2j}(\vec{x},t) |0 \rangle =
{\phi}_{2j}. \label{52}
\end{equation}
The basic points of the large $N$-approximation are as follows.
(i) The equal time two-point correlation functions are
proportional to ${\delta}_{{\alpha},{\beta}}$
$({\alpha},{\beta}=\overline{1,2})$ and ${\delta}_{jl}$
$(j,l=\overline{1,N})$ unless the vacuum expectation values of
$\overline{\chi}_{{\alpha}j}$ are non-zero. Assuming
translational invariance of the vacuum state $|0 \rangle$, we
define the correlation functions
\begin{equation}
\langle 0| \overline{\chi}_{1j}(\vec{x},t) \overline{\chi}_{1l}
(\vec{y},t) |0 \rangle = {\delta}_{jl} G_{1j}(\vec{x}-\vec{y};t),
\label{53}
\end{equation}
\begin{equation}
\langle 0| \overline{\chi}_{2j}(\vec{x},t) \overline{\chi}_{2l}
(\vec{y},t) |0 \rangle = {\phi}_{2j} {\phi}_{2l} + {\delta}_{jl}
G_{2j}(\vec{x}-\vec{y};t), \label{54}
\end{equation}
and
\begin{equation}
\langle 0| \overline{\pi}_{1j}(\vec{x},t) \overline{\chi}_{1l}
(\vec{y},t) |0 \rangle = {\delta}_{jl} D_{1j}(\vec{x}-\vec{y};t),
\label{55}
\end{equation}
\begin{equation}
\langle 0| \overline{\pi}_{2j}(\vec{x},t) \overline{\chi}_{2l}
(\vec{y},t) |0 \rangle = \dot{\phi}_{2j} {\phi}_{2l} +
{\delta}_{jl} D_{2j}(\vec{x}-\vec{y};t), \label{56}
\end{equation}
while
\begin{equation}
\langle 0| \overline{\chi}_{1j}(\vec{x},t) \overline{\chi}_{2l}
(\vec{y},t) |0 \rangle = \langle 0| \overline{\pi}_{1j}(\vec{x},t)
\overline{\chi}_{2l}(\vec{y},t) |0 \rangle =\langle 0|
\overline{\pi}_{2j}(\vec{x},t) \overline{\chi}_{1l} (\vec{y},t) |0
\rangle =0. \label{57}
\end{equation}

(ii) The vacuum values of all fields are the same, i.e.
${\phi}_{1j}={\phi}_1$ and ${\phi}_{2j}={\phi}_2$ for all values
of $j$. As a consequence, ${\mu}_0^2= \overline{\lambda}
{\phi}_1^2$ where $\overline{\lambda} \equiv {\lambda} N$, and the
effective masses coincide, $m_{1j}^2 = m_{2j}^2 \equiv m_{j}^2 =
m^2 + \overline{\lambda}{\phi}_1^2$. The correlation functions
$G_{{\alpha}j}$, $D_{{\alpha}j}$ as well do not depend on the
particular value of $j$. However, in what follows we keep the
index $j$ on all correlation functions, vacuum mean values of
currents and etc, except the ${\phi}$-fields.

(iii) The higher order correlation functions are factorized into
products of two-point correlation functions and vacuum fields. For
example,
\begin{equation}
\langle 0| \overline{\chi}_{2j}(\vec{x},t) \overline{\chi}_{1l}
(\vec{x},t) \overline{\chi}_{1l}(\vec{x},t) |0 \rangle = {\phi}_2
G_{1l}(0;t), \label{58}
\end{equation}
\begin{equation}
\langle 0| \overline{\chi}_{2j}(\vec{x},t) \overline{\chi}_{2l}
(\vec{x},t) \overline{\chi}_{2l}(\vec{x},t) |0 \rangle = {\phi}_2
\Big[ G_{2l}(0;t) + 2{\delta}_{jl} G_{2j}(0;t) + 3{\phi}_2^2
\Big]. \label{59}
\end{equation}

With these points it is straightforward to calculate
${\Delta}_{{\alpha}j}^{(1)}(\vec{k},t)$ and
${\Delta}_{{\alpha}j}^{(2)}(\vec{k},t)$. We start, however, with
the vacuum mean value of the hermitian currents
$\overline{J}_{{\alpha}j}(\vec{x},t)$. Using (\ref{13}) yields for
these currents the following expressions
\begin{equation}
\overline{J}_{1j} = - \sqrt{2{\lambda}} \overline{\chi}_{1j}
\sum_{l=1}^{N} {\mu}_l \overline{\chi}_{1l} - \frac{1}{2} \Big(
\sqrt{2{\lambda}} {\mu}_j + {\lambda} \overline{\chi}_{1j} \Big)
\sum_{l=1}^{N} \Big( \overline{\chi}_{1l}^2 +
\overline{\chi}_{2l}^2 \Big), \label{60}
\end{equation}
\begin{equation}
\overline{J}_{2j} = - \sqrt{2{\lambda}} \overline{\chi}_{2j}
\sum_{l=1}^{N} {\mu}_l \overline{\chi}_{1l} - \frac{\lambda}{2}
\overline{\chi}_{2j} \sum_{l=1}^{N} \Big( \overline{\chi}_{1l}^2 +
\overline{\chi}_{2l}^2 \Big). \label{61}
\end{equation}
Using next the factorization (iii), we obtain
\begin{equation}
\langle \overline{J}_{1j} \rangle = - \frac{\overline{\lambda}}
{\sqrt{2}} {\Omega}_j^2 {\phi}_1, \label{62}
\end{equation}
\begin{equation}
\langle \overline{J}_{2j} \rangle = - \frac{\overline{\lambda}}{2}
\Big[ {\Omega}_j^2 + 3{\phi}_2^2 \Big] {\phi}_2, \label{63}
\end{equation}
where
\begin{equation}
{\Omega}_j^2 \equiv G_{1j}(0;t) + G_{2j}(0;t) = \frac{1}{L^3}
\sum_{{\alpha}=1}^2 \sum_{\vec{k}} \langle 0|
\overline{\chi}_{{\alpha}j}^{\dagger}(\vec{k},t)
\overline{\chi}_{{\alpha}j}(\vec{k},t) |0 \rangle. \label{64}
\end{equation}
As seen from Eq.(\ref{63}), $\langle \overline{J}_{2j} \rangle =0$
for ${\phi}_2=0$, i.e. the condition (\ref{16}) is fulfilled. We
come therefore to the important conclusion: in the large-$N$
approximation the ${\phi}$-fields can be taken neutral at all
times. Henceforth we put ${\phi}_2=0$.

Substituting Eq.(\ref{62}) into Eq.(\ref{14}), we rewrite the
vacuum mean field equation as
\begin{equation}
\ddot{\phi}_{1j} + \Big( m_j^2 + \frac{\overline{\lambda}}{2}
{\Omega}_j^2 \Big) {\phi}_{1j} =0, \label{65}
\end{equation}
so that the higher order effects contribution to the effective
masses is
\begin{equation}
M_j^2 = \frac{\overline{\lambda}}{2} {\Omega}_j^2 . \label{66}
\end{equation}

We turn now to the calculation of ${\Delta}_{{\alpha}j}^{(1)}
(\vec{k},t)$ and ${\Delta}_{{\alpha}j}^{(2)}(\vec{k},t)$. In the
large-$N$ approximation, due to the point $(i)$ the interaction of
components with different values of $j$ does not contribute to the
vacuum expectation values, and
${\Delta}_{{\alpha}j}^{(1)}(\vec{k},t)$ and
${\Delta}_{{\alpha}j}^{(2)}(\vec{k},t)$ take the form
\begin{eqnarray}
{\Delta}_{{\alpha}j}^{(1)}(\vec{k},t) & = &
\frac{1}{\tilde{\omega}_{{\alpha}j}} {\rm Re}\langle 0|
\overline{\pi}_{{\alpha}j}(\vec{k},t)
\overline{J}_{{\alpha}j}(\vec{k},t) |0 \rangle +
\frac{\overline{\lambda}}{2\tilde{\omega}_{{\alpha}j}}
{\Omega}_j^2 \cdot {\rm Re} D_{{\alpha}j}(\vec{k},t),\\
{\Delta}_{{\alpha}j}^{(2)}(\vec{k},t) & = & {\rm Re}\langle 0|
\overline{\chi}_{{\alpha}j}^{\dagger}(\vec{k},t)
\overline{J}_{{\alpha}j}(\vec{k},t)|0 \rangle +
\frac{\overline{\lambda}}{2} {\Omega}_j^2 \cdot G_{{\alpha}j}
(\vec{k},t), \label{67-68}
\end{eqnarray}
where $D_{{\alpha}j}(\vec{k},t)$ and $G_{{\alpha}j}(\vec{k},t)$
are the Fourier transforms of $D_{{\alpha}j}(\vec{x}-\vec{y};t)$
and $G_{{\alpha}j}(\vec{x}-\vec{y};t)$, respectively.

Evaluating next the self-interaction contribution by making use of
(iii), we obtain
\begin{eqnarray}
{\rm Re}\langle 0|\overline{\pi}_{{\alpha}j}(\vec{k},t)
\overline{J}_{{\alpha}j}(\vec{k},t) |0\rangle & = & -
\frac{\overline{\lambda}}{2} {\Omega}_j^2 \cdot {\rm
Re}D_{{\alpha}j}(\vec{k},t), \\
{\rm Re}\langle 0|\overline{\chi}_{{\alpha}j}^{\dagger}(\vec{k},t)
\overline{J}_{{\alpha}j}(\vec{k},t) |0\rangle & = & -
\frac{\overline{\lambda}}{2} {\Omega}_j^2 \cdot
G_{{\alpha}j}(\vec{k},t), \label{69-70}
\end{eqnarray}
leading finally to
\begin{equation}
{\Delta}_{{\alpha}j}^{(1)}(\vec{k},t)={\Delta}_{{\alpha}j}^{(2)}(\vec{k},t)=0.
\label{71}
\end{equation}

In the same way we calculate the vacuum expectation value in the
r.h.s. of (\ref{44}) and obtain that
\begin{equation}
\dot{\cal N}_{{\alpha}j,-}(\vec{k},t) =0, \label{72}
\end{equation}
i.e. in the large-$N$ approximation the particle production is
symmetric in the momentum space for all times.

With ${\Delta}^{(1)}_{{\alpha}j}={\Delta}^{(2)}_{{\alpha}j}=0$,
Eq.(\ref{50}) is solved by
\begin{equation}
C_{{\alpha}j}(\vec{k},t)= e^{2i{\Theta}_{{\alpha}j}} \int_{t_0}^t
dt^{\prime} \frac{\dot{\tilde{\omega}}_{{\alpha}j}}
{2\tilde{\omega}_{{\alpha}j}}(\vec{k},t^{\prime}) \Big( 1 + 2{\cal
N}_{{\alpha}j,+}(\vec{k},t^{\prime}) \Big) \cdot
e^{2i(\tilde{\Theta}_{{\alpha}j}(\vec{k},t^{\prime}) -
\tilde{\Theta}_{{\alpha}j}(\vec{k},t))}, \label{73}
\end{equation}
where
\begin{equation}
\tilde{\Theta}_{{\alpha}j}(\vec{k},t) \equiv \int_{t_0}^t
dt^{\prime} \tilde{\omega}_{{\alpha}j}(\vec{k},t^{\prime}).
\label{74}
\end{equation}
Substituting (\ref{73}) into (\ref{46}), we obtain the following closed
equation for ${\cal N}_{{\alpha}j,+}(\vec{k},t)$:
\begin{equation}
\dot{\cal N}_{{\alpha}j,+}(\vec{k},t) = \frac{1}{2}
\tilde{W}_{{\alpha}j}(\vec{k},t) \int_{t_0}^t dt^{\prime}
\tilde{W}_{{\alpha}j}(\vec{k},t^{\prime}) \cdot \Big( 1 + 2{\cal
N}_{{\alpha}j,+}(\vec{k},t^{\prime}) \Big) \cos\Big[
\tilde{x}_{{\alpha}j} (\vec{k};t^{\prime},t) \Big], \label{75}
\end{equation}
where
\begin{equation}
\tilde{W}_{{\alpha}j} \equiv
\frac{\dot{\tilde{\omega}}_{{\alpha}j}}{\tilde{\omega}_{{\alpha}j}}
\label{76}
\end{equation}
are kinetic amplitudes, while
\begin{equation}
\tilde{x}_{{\alpha}j}(\vec{k};t^{\prime},t) \equiv
2\Big[ \tilde{\Theta}_{{\alpha}j}(\vec{k},t^{\prime}) -
\tilde{\Theta}_{{\alpha}j}(\vec{k},t) \Big]
\label{77}
\end{equation}
is the phase difference.

Eq.(\ref{75}) is a complete quantum kinetic equation in the
large-$N$ approximation with all higher order effects included.
It determines the time evolution of the number of particles of a
fixed $j$ and a fixed momentum $\vec{k}$.

The background ${\phi}$-fields are given by solution of the
non-linear equation (\ref{65}), while ${\Omega}_j^2$ must be
determined self-consistently from
\begin{equation}
{\Omega}_j^2 = \frac{1}{L^3} \sum_{{\alpha}=1}^{2} \sum_{\vec{k}}
\frac{1}{2\tilde{\omega}_{{\alpha}j}} \Big[ \Big( 1 + 2{\cal
N}_{{\alpha}j,+}(\vec{k},t) \Big) + 2R_{{\alpha}j}(\vec{k},t)
\Big] \label{78}
\end{equation}
with
\begin{equation}
R_{{\alpha}j}(\vec{k},t) = \frac{1}{2} \int_{t_0}^t dt^{\prime}
\tilde{W}_{{\alpha}j}(\vec{k},t^{\prime}) \Big(1+2{\cal
N}_{{\alpha}j,+} (\vec{k},t^{\prime}) \Big)  \cos\Big[
\tilde{x}_{{\alpha}j} (\vec{k};t^{\prime},t) \Big]. \label{79}
\end{equation}
We have therefore a system of three coupled equations.

The kinetic equation (\ref{75}) has the following important
features: (i) The source term, i.e. its r.h.s. is non-Markovian.
Therefore calculating the particle density at any given instant
requires a complete knowledge of the history of the production
process. In addition, the integrand is a non-local function of
time, which apparent in the coherent phase oscillation term $\cos
\Big[ \tilde{x}_{{\alpha}j}(\vec{k};t^{\prime},t) \Big]$; (ii) The
background ${\phi}$-fields do not contribute to the kinetic
equation directly, but only via the frequencies of the quantum
fluctuations, as evident in Eqs.(\ref{76}) and (\ref{77}); (iii)
The production rate is affected by the produced particles'
statistics, as seen in the statistical factor $(1+2{\cal
N}_{{\alpha}j,+}(\vec{k},t))$. In the case of fermions, this
factor would be $(1-2{\cal N}_{{\alpha}j,+} (\vec{k},t))$.

The kinetic amplitudes  $\tilde{W}_{{\alpha}j}(\vec{k},t)$ are
decomposed as
\begin{equation}
\tilde{W}_{{\alpha}j}=W_{{\alpha}j} + W_{{\alpha}j}^{\Omega},
\label{80}
\end{equation}
where
\begin{equation}
W_{{\alpha}j} \equiv \frac{\dot{\omega}_{{\alpha}j}}
{{\omega}_{{\alpha}j}}
\label{81}
\end{equation}
and
\begin{equation}
W_{{\alpha}j}^{\Omega} \equiv \frac{\overline{\lambda}}{2}
\frac{{\Omega}_j}{{\omega}_{{\alpha}j}^2} \Big( \dot{\Omega}_j -
{\Omega}_j W_{{\alpha}j} \Big) \Big( 1 +
\frac{\overline{\lambda}}{2}
\frac{{\Omega}_j^2}{{\omega}_{{\alpha}j}^2} \Big)^{-1}. \label{82}
\end{equation}
The amplitudes $W_{{\alpha}j}(\vec{k},t)$
account for the increase in the particle number caused by the
transition of energy from the background ${\phi}$-fields to
fluctuations. As soon as the energy transition stops, the amplitudes
$W_{{\alpha}j}(\vec{k},t)$ vanish. If, for example, the time evolution
of the ${\phi}$-fields starts
at $t=t_0$ and ends at $t=t_1$, then $W_{{\alpha}j}(\vec{k},t)=0$
for all $t>t_1$.

In contrast with $W_{{\alpha}j}(\vec{k},t)$, the new amplitudes
$W_{{\alpha}j}^{\Omega}(\vec{k},t)$ account for
the change in the particle number density caused by the
interaction of the particles produced, i.e. by their collisions, and
do not vanish when the particle production by the background fields
stops. The fluctuations interact with each other in two ways, directly and
via the ${\phi}$-fields. For $t>t_1$ , the time-dependence of
the particle number density is only governed by collisions.

Collisions also modify the expression for the dynamical phase. If
collision effects are omitted, $\tilde{\Theta}_{{\alpha}j}$
reduces to the adiabatic phase
\begin{equation}
{\Theta}_{{\alpha}j}^{ad}(\vec{k},t) \equiv
\int_{t_0}^t dt^{\prime} {\omega}_{{\alpha}j}(\vec{k},t^{\prime})
\label{83}
\end{equation}
that takes the form ${\omega}_k^{(0)}(t-t_0)$ in the in-limit.
With collisions taken into account, the dynamical phase is no
longer the adiabatic one and the connection with the in-picture no
longer exists.

The non-Markovian kinetic equation in the collisionless limit can
be simply obtained from (\ref{75}) by replacing
$\tilde{W}_{{\alpha}j}$ with $W_{{\alpha}j}$ and
$\tilde{\Theta}_{{\alpha}j}$ with ${\Theta}_{{\alpha}j}^{ad}$. The
kinetic equations in both cases, i.e. with and without collisions,
have therefore the same non-Markovian structure. This is a result
of the large-$N$ approximation. However, collisions introduce
non-Markovian effects in addition to those already present in the
case without collisions. Even in the "collision" regime for
$t>t_1$, the production rate is affected by the background fields
induced particle production process from $t_0$ to $t_1$ .

%%%%%%%%%%%%%%%%%%%%%%%%%%%%%%%%%%%%%%%%%%%%%%%%%%%%%%%%%%%%%%%%%
\subsection{Renormalization}
%%%%%%%%%%%%%%%%%%%%%%%%%%%%%%%%%%%%%%%%%%%%%%%%%%%%%%%%%%%%%%%%%%

All the divergences in our approach come from the expression for $
\tilde{\omega}_{{\alpha}j}(t)$:
\begin{equation}
\tilde{\omega}_{{\alpha}j}^2 = \vec{k}^2 + m_{{\alpha}j}^2(t) +
\frac{\overline{\lambda}}{2} {\Omega}_j^2(t), \label{84}
\end{equation}
the sum over $\vec{k}$ in the definition of ${\Omega}_j^2$ $\Big[
{\rm Eq.}(\ref{78}) \Big]$ having both quadratic and logarithmic
divergences which have to be absorbed by mass and coupling
constant renormalization \cite{CM1987} . The subtraction of
${\Omega}_j^2(t_0)$ from ${\Omega}_j^2(t)$ removes the quadratic
divergence. With ${\cal
N}_{{\alpha}j,+}(\vec{k},t_0)=R_{{\alpha}j}(\vec{k},t_0)=0$, we
rewrite $\tilde{\omega}_{{\alpha}j}^2(t)$ as
\begin{equation}
\tilde{\omega}_{{\alpha}j}^2(t) = {\omega}_R^2 + S_j(t),
\label{85}
\end{equation}
where
\begin{equation}
{\omega}_R^2 \equiv \vec{k}^2 + m_R^2, \label{86}
\end{equation}
$m_R^2$ being the renormalized mass squared,
\begin{equation}
m_R^2 \equiv m^2 + \overline{\lambda} {\phi}_1^2(t_0) +
\frac{\overline{\lambda}}{2} {\Omega}_j^2(t_0), \label{87}
\end{equation}
and
\begin{equation}
S_j(t) \equiv \overline{\lambda} \Big[ {\phi}_1^2(t) -
{\phi}_1^2(t_0) \Big] + \frac{\overline{\lambda}}{2} \Big[
{\Omega}_j^2(t) - {\Omega}_j^2(t_0) \Big], \label{88}
\end{equation}
$S_j(t_0)=0$.

The remaining logarithmic divergence is removed by the coupling
constant renormalization. The bare coupling constant and the
renormalized one are related by the geometric series
\begin{equation}
\overline{\lambda}= \frac{{\lambda}_R}{1 - {\lambda}_R
{\delta}{\lambda}}, \label{89}
\end{equation}
where
\begin{equation}
{\delta}{\lambda} \equiv \frac{1}{L^3} \sum_{\vec{k}}
\frac{1}{4{\omega}_R^3(\vec{k})}. \label{90}
\end{equation}
Multiplying both sides of (\ref{88}) by $(1-{\lambda}_R
{\delta}{\lambda} )$, we obtain the finite equation
\begin{equation}
S_j(t)={\lambda}_R \Big[ {\phi}_1^2(t) - {\phi}_1^2(t_0) \Big] +
\frac{{\lambda}_R}{2} \Big[ {\Omega}_j^2(t) - {\Omega}_j^2(t_0) +
2{\delta}{\lambda} \cdot S_j(t) \Big]. \label{91}
\end{equation}

If we expand ${\Omega}_j^2(t)$ and $S_j(t)$ in a power series in
${\lambda}_R$,
\begin{equation}
{\Omega}_j^2(t) = {\Omega}_{j(0)}^2(t) + \sum_{n=1}^{\infty}
{\lambda}_R^n \cdot {\Omega}_{j(n)}^2(t), \label{92}
\end{equation}
\begin{equation}
S_j(t) = \sum_{n=1}^{\infty} {\lambda}_R^n \cdot S_{j(n)}(t),
\label{93}
\end{equation}
and choose the initial conditions for ${\Omega}_j^2(t)$ as
\begin{equation}
{\Omega}_j^2(t_0) = {\Omega}_{j(0)}^2(t_0), \label{94}
\end{equation}
\begin{equation}
{\Omega}_{j(k)}^2(t_0) =0 \hspace{5 mm} {\rm for} \hspace{5 mm} k
\ge 1, \label{95}
\end{equation}
then (\ref{91}) reduces to the following set of equations
\[
S_{j(1)}(t) = \Big[ {\phi}_1^2(t) - {\phi}_1^2(t_0) \Big] +
\frac{1}{2} \Big[ {\Omega}_{j(0)}^2(t) - {\Omega}_{j(0)}^2(t_0)
\Big],
\]
\begin{equation}
S_{j(2)}(t) = \frac{1}{2} {\Omega}_{j(1)}^2(t) + {\delta}{\lambda}
\cdot S_{j(1)}(t), \label{96}
\end{equation}
\[
S_{j(3)}(t) = \frac{1}{2} {\Omega}_{j(2)}^2(t) + {\delta}{\lambda}
\cdot S_{j(2)}(t), . . . ,
\]
i.e. in the large-$N$ approximation we need only to calculate in
the second order to identify the counterterms \cite{CM1987}.

The renormalized vacuum mean field equation takes the form
\begin{equation}
\ddot{\phi}_{1j} + \Big( m_R^2 + S_j(t) \Big) {\phi}_{1j} =0.
\label{97}
\end{equation}
Expanding the amplitudes $\tilde{W}_{{\alpha}j}$ and the dynamical
phases $\tilde{\Theta}_{{\alpha}j}$ in power series in
${\lambda}_R$, it is also possible to deduce from Eqs.(\ref{75})
and (\ref{78})-(\ref{79}) a renormalized kinetic equation in a
fixed order of perturbations as well as to develop a perturbation
theory technique for the particle production problem. This work is
in progress and will be reported elsewhere.

%%%%%%%%%%%%%%%%%%%%%%%%%%%%%%%%%%%%%%%%%%%%%%%%%%%%%%%%%%%%%%%%%%%%%%%%
\section{Discussion}
%%%%%%%%%%%%%%%%%%%%%%%%%%%%%%%%%%%%%%%%%%%%%%%%%%%%%%%%%%%%%%%%%%%%%%%%

1. For the model of a self-interacting complex $N$-component
scalar field, we have derived in the large-$N$ approximation a
quantum kinetic equation with a non-Markovian source term. This
equation determines the momentum distribution of the standard,
non-tachyonic modes produced in quantum fluctuations of the scalar
field around its vacuum mean value.

Our equation is complete in the sense that it includes all orders
of the back-reaction and collision effects. Its form is compact
and appropriate for the numerical study. We have obtained an
explicit expression for the kinetic amplitudes related to
collisions. These amplitudes account for the change in the
particle number density in collisions in both "energy transition"
and "collision" regimes.

Our method of derivation of kinetic equations is not restricted to
the model discussed. It can be applied to any self-interacting
field theory, providing the approximation scheme to treat
non-diagonal interactions is specified. We have shown that in the
large-$N$ approximation the functions
${\Delta}_{{\alpha}j}^{(1)}$, ${\Delta}_{{\alpha}j}^{(2)}$ vanish,
so that the source term with higher order effects included has the
same non-Markovian structure as the one with higher order effects
omitted. However, beyond the large-$N$ approximation, the
functions ${\Delta}_{{\alpha}j}^{(1)}$,
${\Delta}_{{\alpha}j}^{(2)}$ are expected to modify the structure
of the source term as well.

2. For the particle production induced by an external electric
field, the non-Markovian effects are known to be important when
the field is strong \cite{S1998}. For weak fields, non-Markovian
effects disappear, and the Markovian limit of the kinetic
equation, which is defined by the neglect of memory effects in the
source term, can be successfully employed. If the external field
is switched off, then the particle production stops.

For the self-interacting fields, the situation is completely
different. The self-interaction takes place at all times and at
all values of fields. It introduces memory effects, which appear
even at lower orders of perturbations and can not be therefore
neglected in any sensible limit or regime. Once the
self-interaction is taken into consideration, the non-Markovian
character of the source term is unavoidable. Even if the
transition of energy from the background fields to fluctuations
stops, the particle production process continues, being influenced
essentially by the background field evolution history.

%%%%%%%%%%%%%%%%%%%%%%%%%%%%%%%%%%%%%%%%%%%%%%%%%%%%%%%%%%%%%%%%%%%%%%%%%%%%


\begin{thebibliography}{99}
\bibitem{RM1996}
J.~Rau and B.~M\"uller,
Phys.\ Repts. {\bf 272} (1996) 1.
\bibitem{KME1998}
Y.~Kluger, E.~Mottola, and J.M.~Eisenberg, Phys.\ Rev. {\bf D 58}
(1998) 125015.
\bibitem{S1998}
S.~Schmidt et al., Int.\ J.\ Mod.\ Phys. {\bf E 7} (1998) 709;
Phys.\ Rev. {\bf D 59} (1999) 094005; J.C.R.~Bloch et al., Phys.\
Rev. {\bf D 60} (1999) 116011; C.D.~Roberts and S.M.~Schmidt,
Progr.\ Part.\ Nucl.\ Phys. {\bf 45} (2000) pp. S1-S103.
\bibitem{GWR1994}
C.~Greiner, K.~Wagner, and P.-G. Reinhard, Phys.\ Rev. {\bf C 49}
(1994) 1693; H.~Heiselberg and X.~Wang, Phys.\ Rev. {\bf C 53}
(1996) 1892; T.S.~Biro and C.~Greiner, Phys.\ Rev.\ Lett. {\bf 79}
(1997) 3138; W.M.~Alberico, A.~Lavagno, and P.~Quarati, Eur.\
Phys.\ J. {\bf C 12} (2000) 499; O.V.~Utyuzh, G.~Wilk, and
Z.~Wlodarczyk, J.\ Phys. {\bf G 26} (2000) L39.
\bibitem{AYGS1998}
S.~Ayik, M.~Belkacem, and A.~Bonasera, Phys.\ Rev. {\bf C 51}
(1995) 611; S.~Ayik et al., Phys.\ Rev. {\bf C 58} (1998) 1594.
\bibitem{CDG1994}
M.~Colonna, M.~Di Toro, and A.~Guarnera, Nucl.\ Phys. {\bf A 580}
(1994) 312.
\bibitem{FMW1998}
U.~Fuhrmann, K.~Morawetz, and R.Walke, Phys.\ Rev. {\bf C 58}
(1998) 1473.
\bibitem{G1996}
K.~Geiger, Phys.\ Rev. {\bf D 54} (1996) 949; {\bf D 56} (1996)
2665; S.~Mrowczynski, Fiz.\ Elem.\ Chast.\ At.\ Yadra {\bf 30}
(1999) 954 [ Phys.\ Part.\ Nuclei {\bf 30} (1999) 419 ]; S.A.~Bass
et al., Progr.\ Part.\ Nucl.\ Phys. {\bf 41} (1998) 225.
\bibitem{B1999}
D.~Bodeker, Nucl.\ Phys. {\bf B 566} (2000) 402; P.~Arnold,
D.T.~Son, and L.G.~Yaffe, Phys.\ Rev. {\bf D 59} (1999) 105020;
J.-P.~Blaizot and E.~Iancu, Nucl.\ Phys. {\bf B 557} (1999) 183;
{\bf B 570} (2000) 326; D.F.~Litim and C.~Manuel, Nucl.\ Phys.
{\bf B 562} (1999) 237.
\bibitem{BLL1996}
D.~Boyanovsky, I.D.~Lawrie, and D.-S.~Lee, Phys.\ Rev. {\bf D 54}
(1996) 4013; I.D.~Lawrie and D.B.~McKernan, Phys.\ Rev. {\bf D 55}
(1997) 2290; F.T.~Brandt, J.~Frenkel, and A.~Guerra, Int.\ J.\
Mod.\ Phys. {A 13} (1998) 4281.
\bibitem{BVW2000}
D.~Boyanovsky, H.J.~de Vega, and S.-Y.~Wang, Phys.\ Rev. {\bf D
61} (2000) 065006.
\bibitem{B2002}
D.B.~Blaschke et al., Phys.\ Rev. {\bf D 65} (2002) 054039.
\bibitem{S2003}
F.M.~Saradzhev, Phys.\ Lett. {\bf B 558} (2003) 103.
\bibitem{K1989}
J.~Kapusta, {\it Finite Temperature Field Theory} (Cambridge
Monographs on Mathematical Physics, 1989).
\bibitem{CJT1974}
J.~Cornwall, R.~Jackiw, and E.~Tomboulis, Phys.\ Rev. {\bf D 10}
(1974) 2428; S.~Coleman, R.~Jackiw, and H.D.~Politzer, Phys.\ Rev.
{\bf D 10} (1974) 2491; D.J.~Gross and A.~Neveu, Phys.\ Rev. {\bf
D 10} (1975) 3235; F.~Cooper, G.S.~Guralnik, and S.H.~Kasdan,
Phys.\ Rev. {\bf D 14} (1976) 1607.
\bibitem{BG1980}
T.~Barnes and G.~Ghandour, Phys.\ Rev. {\bf D 22} (1980) 924;
W.A.~Bardeen and M.~Moshe, Phys.\ Rev. {\bf D 28} (1983) 1372;
C.~Bender and F.~Cooper, Ann.\ Phys. (N.Y.) {\bf 160} (1985) 323.
\bibitem{M1985}
G.F.~Mazenko, Phys.\ Rev.\ Lett. {\bf 54} (1985) 2163; Phys.\ Rev.
{\bf D 34} (1986) 2223.
\bibitem{CM1987}
F.~Cooper and E.~Mottola, Phys.\ Rev. {\bf D 36} (1987) 3114;
Phys.\ Rev. {\bf D 40} (1989) 456.
\end{thebibliography}
\end{document}